\documentclass{aastex63}
\usepackage{amsmath}
\usepackage[]{hyperref}
\usepackage{longtable}

\hypersetup{linkcolor=red,citecolor=blue,filecolor=cyan,urlcolor=magenta}

\received{2019 November 26}
\accepted{2019 December 22}
\published{2020 January 16}

\begin{document}

\title{The nearby luminous transient AT2018cow: a magnetar formed in a sub-relativistically expanding non-jetted explosion}

\correspondingauthor{P. Mohan}
\email{pmohan@shao.ac.cn}
\correspondingauthor{Tao An}
\email{antao@shao.ac.cn}

\author{P. Mohan}
\affiliation{Shanghai Astronomical Observatory, 80 Nandan Road, Shanghai 200030, China}

\author{T. An}
\affiliation{Shanghai Astronomical Observatory, 80 Nandan Road, Shanghai 200030, China}
\affiliation{Key Laboratory of Radio Astronomy, Chinese Academy of Sciences, 210008 Nanjing, China}

\author{J. Yang}
\affiliation{Department of Space, Earth and Environment, Chalmers University of Technology\\
Onsala Space Observatory, SE-439 92 Onsala, Sweden}
\affiliation{Shanghai Astronomical Observatory, 80 Nandan Road, Shanghai 200030, China}
 
\begin{abstract}
The fast-rising blue optical transient AT2018cow indicated unusual early phase characteristics unlike relatively better studied explosive transients. Its afterglow may be produced by either a relativistically beamed (jetted) or intrinsically luminous (non-jetted) ejecta and carries observational signatures of the progenitor and environment. High resolution monitoring can distinguish between these scenarios and clarify the progenitor nature. We present very long baseline interferometry (VLBI) observations of AT2018cow at 5 GHz involving 21 radio telescopes from the European VLBI Network with five sessions spanning $\approx$ 1 year. With an astrometric precision up to 25 micro-arcseconds ($\mu$as) per epoch, the rapidly fading compact mas scale source is found to be non-jetted with a proper motion of $\leq$ 0.15 mas yr$^{-1}$ (0.14 $c$). This and a dense (number density $\approx 10^4 - 10^5$ cm$^{-3}$) magnetized environment (magnetic field strength $\geq$ 0.84 G) are characteristic of a newly formed magnetar driven central engine, originating in the successful explosion of a low-mass star.
\end{abstract}


\section{Introduction} \label{sec:intro}

Transient astrophysical events are increasingly detected by optical and high energy (X-rays, gamma-rays) survey telescopes \cite[e.g.][]{2014ApJ...794...23D,2019NatAs...3..697I}. Owing to advances in data collection and processing, their identification is near real-time with follow-up monitoring observations being triggered within a few hours. These events are generally indicative of cataclysmic cosmic explosions. They involve the core-collapse of a massive star or merger of two stars (e.g. supernovae and gamma-ray bursts), or the tidal disruption of a star by a massive black hole. Fast-rising blue optical transients (FBOTs) are characterized by a rapid rise to a peak (timescale of $\leq$ 10 days, luminosity of $10^{43} - 10^{44}$ erg s$^{-1}$) or a time above half maximum of the luminosity $t_{1/2} \leq$ 12 days and a strongly blue colour ($g-r<-0.2$) near the peak following which is an exponential decline within 30 days \cite[e.g.][]{2019NatAs...3..697I,2018MNRAS.481..894P}. Their origin and progenitors are poorly understood mainly due to the current lack of dedicated multi-wavelength monitoring observations. The FBOT AT2018cow was identified in the Asteroid Terrestrial-impact Last Alert System (ATLAS) survey on 16 June 2018 \citep{2018ATel11727....1S}, hosted in the dwarf spiral galaxy CGCG 137-068 at a redshift of 0.014 \citep{2018ApJ...865L...3P}. It shows peculiar properties: the fastest rise with a timescale $<$ 3.3 days, a $t_{1/2}$ of $<$ 1.7 days  (during the rising phase), a high peak luminosity of $(1.7 - 4.0)\times 10^{44}$ erg s$^{-1}$ (exceeding that of super-luminous supernovae), and an unusual featureless non-evolving early-phase spectrum \citep{2018ApJ...865L...3P} followed by a rapid decline in the light curve \citep{2018ApJ...865L...3P,2019MNRAS.484.1031P,2019ApJ...872...18M}. This made it one of the most extreme amongst the current sample of FBOTs and thus presented a unique opportunity for the continued monitoring to help clarify its origin and physical properties. 

Multi-wavelength studies of AT2018cow mostly probed the early optically thick rising phase of spectral evolution \citep{2018ApJ...865L...3P,2018MNRAS.480L.146R,2019MNRAS.484.1031P,2019MNRAS.487.2505K,2019ApJ...871...73H,2019ApJ...872...18M}. Optical spectroscopy indicates a peak blackbody temperature  of $\approx$ 27000 K, ejecta mass of 0.1 - 0.4 $M_\odot$ and a transition from an initial featureless spectrum to the appearance of distinct broad H and He emission lines only after 15 days from the discovery \citep{2018ApJ...865L...3P,2019MNRAS.484.1031P,2019ApJ...872...18M}. The early X-ray flux was variable over day timescales, suggestive of a shock interacting with a non-uniform surrounding medium \citep{2019ApJ...872...18M}. The early phase self-absorbed radio emission is indicative of an energetic interaction with a dense medium (number density of $3\times 10^4 - 4\times 10^5$ cm$^{-3}$) \citep{2019ApJ...871...73H}. A powering by the radioactive decay of $^{56}$Ni, expected in core-collapse supernovae was ruled out based on an nonphysical requirement on the Ni mass ($\approx$ 3 $M_\odot$) in comparison to the inferred ejecta mass of $\approx$ 0.1 - 0.4 $M_\odot$ and the inability to account for the light curve decay \citep{2018ApJ...865L...3P}. Powering by the tidal disruption of a white dwarf star by an intermediate mass black hole ($10^4 M_\odot$) \citep{2019MNRAS.484.1031P,2019MNRAS.487.2505K} was disfavored owing to a dense \citep{2019ApJ...872...18M,2019ApJ...871...73H} highly magnetized \citep{2019ApJ...878L..25H} environment inferred in the immediate vicinity of the transient, which may be challenging to naturally develop in this scenario and require a pre-existing reservoir \citep{2019ApJ...872...18M}. The absence of excitation H and He narrow lines following the transition from a high velocity expansion ($\geq 0.1 c$) to a slower one ($\approx$ 0.02 $c$) seem to disfavor an origin in a failed regular supernovae of a giant star (with a consequent direct collapse to a black hole) \citep{2019MNRAS.484.1031P}. 

These multi-wavelength studies find evidence for a central engine consisting of a compact object (newly formed stellar mass black hole or neutron star/magnetar) powering the early phase of the source through accretion – jet/outflow production and sustenance \citep{2019ApJ...872...18M,2019ApJ...871...73H,2019ApJ...878L..25H,2019MNRAS.487.5618L}. The polar directed outflow \citep{2019ApJ...872...18M,2019MNRAS.484.4972S} interacts with the dense, stratified medium \citep{2018MNRAS.480L.146R} producing the observed flux density evolution and variability \citep{2018MNRAS.480L.146R,2019ApJ...871...73H}. There is then the possibility of discovering a collimated relativistic jet with Doppler boosted emission if it can be sustained by the central engine. However, a recent VLBI study of AT2018cow (spanning up to 287 days after discovery) at 22 GHz and 8.4 GHz finds a symmetric expansion with proper motion constrained to $\leq$ 0.51 $c$ and suggests that the jet may not be long lived \citep{2019MNRAS.tmp.2832B}, thus requiring confirmation. 

The luminous afterglow emission in AT2018cow carries observational signatures of the progenitor and environment. High resolution monitoring can help pin down the transient origin through inferred physical properties and stringent constraints on proper motion. This can be accomplished with very long baseline interferometry (VLBI) imaging, which has been used in the past to understand  the structure  and environment of gamma-ray bursts \citep{2004ApJ...609L...1T,2019Sci...363..968G} and tidal disruption events \citep{2016MNRAS.462L..66Y}. The obtained information from VLBI observations is crucial in localizing the afterglow and inferring if it is jetted/collimated or non-jetted/wide-angled, constraining the explosion energy, number density and distribution of the surrounding medium. 

We report our imaging observations of AT2018cow over five sessions (between $t_d$ of 94 days and 355 days after discovery) using the European VLBI Network (EVN), conducted at 5 GHz to ensure a detection over a long duration. As these observations cover the turnover and late optically thin spectral evolution phase, they provide unique information to discern the source nature, especially as the source faded rapidly rendering it challenging for other multi-wavelength instruments to capture. With the cosmological parameters H$_{0}$ = 70 km s$^{-1}$ Mpc$^{-1}$, $\Omega_{m} = 0.27$, $\Omega_{\Lambda} = 0.73$, an angular size of 1 mas corresponds to a projected size of 0.286 pc at $z = 0.014$ \citep{2006PASP..118.1711W} and a proper motion speed of 1 mas yr$^{-1}$ corresponds to 0.93 $c$. 

\section{Observations and data processing}

\subsection{Observational setup and calibrators}
We observed AT2018cow with the EVN at 1.67 GHz in the first experiment and at 5 GHz in the later four experiments. A total of 21 radio telescopes spreading Africa, Asia, Europe, and US participated in the observations. These include: Ar (Arecibo 300m, USA), Bd (Badary 32m, Russia),  Cm (Cambridge 32m, UK), De (Defford 25m, UK), Ef (Effelsberg 100m, Germany), Hh (Hartebeesthoek 26m, South Africa), Ib (Irbene 32m, Latvia), Ir (Irbene 16m, Latvia), Jb2 (Lovell 76m, UK), Km (Kunming 40m, China), Kn (Knockin 25m, UK), Mc (Medicina  25m, Italy), On85 (Onsala 25m, Sweden), Sr (Sardinia 64m, Italy), Sv (Svetloe 32m, Russia), T6 (Tianma 65m, China), Tr (Torun 32m, Poland), Ur (Urumqi 26m, China), Wb (Westerbork 25m, Netherlands), Ys (Yebes 40m, Spain), Zc (Zelenchukskaya 32m, Russia). The experiment setup details are summarized in \autoref{table1} and include the EVN project code, observation date, observation frequency, time duration, bandwidth and participating telescopes. The participation of Chinese and Russian telescopes in Sessions 2, 4 and 5 significantly increases the East-West direction resolution to the one mas level. 

All observations were carried out in the phase-referencing mode. J1619+2247 ($\approx$ 0.6 Jy at 5 GHz, 55 arcmin away from AT2018cow) was used as the primary calibrator. Its coordinate is $\alpha$ (J2000) = 16$^h$19$^m$14$^{s}$.8245991 and $\delta$(J2000) = 22$^\circ$47$^{'}$47$^{''}$.851082 in the source catalogue of GSFC 2015a from the Goddard Space Flight Centre (GSFC) VLBI group. We observed J1619+2247 and AT2018cow with a cycle of 220 seconds (40 sec on J1619+2247 and 180 sec on AT2018cow).  In order to further improve the astrometric precision and verify the positional stability of the reference source, we observed two additional sources for a two-minute scan per half hour. They are weaker but closer: NVSS J161541.6+221629 (4.3 arcmin away from AT2018cow, named R1 hereafter), and NVSS J161640.8+221856 (angular separation of 9.8 arcmin, named R2 hereafter). A bright quasar J1642+3948 ($\approx$ 6 Jy) was observed as the fringe finder.

\subsection{Data processing}
The correlation was done by the EVN software correlator SFXC \citep{2015ExA....39..259K} at Joint Institute for VLBI ERIC (JIVE) using the typical correlation parameters (integration time: 1 sec,  frequency resolution: 1 MHz). In the eEVN observations of RY007A (Session 1) and EY033A (Session 3), the data were transferred via broad-band optical fiber and correlated in real time. The real-time correlation and the rapid distribution of the first session correlated data played a key role in ensuring the successful detection of AT2018cow and verifying the suitability of the two nearby calibrators R1 and R2. 

The visibility data were calibrated using the software package Astronomical Image Processing System (AIPS) \citep{2003ASSL..285..109G} of National Radio Astronomy Observatory (NRAO). The side channels with low amplitude were dropped out in loading the data. The AIPS task ACCOR was performed to re-normalize the visibility amplitude. A-priori amplitude calibration was performed with the system temperatures and the antenna gain curves provided by each station. In case that some telescope monitoring data were missing, nominal values of the system equivalent flux density in the EVN status table were used instead. The ionospheric dispersive delays were corrected using a map of total electron content provided by Global Positioning System (GPS) satellite observations. Phase errors due to antenna parallactic angle variations were removed. After a manual phase calibration and bandpass calibration was carried out using the fringe finder, the global fringe-fitting was performed on the phase-referencing calibrators.   

We imaged the calibrator J1619+2247 in DIFMAP software package \citep{1994BAAS...26..987S}. It shows a one-side core-jet structure with a total flux density of 0.61$\pm$0.03 Jy. Its core, i.e., the optically thick jet base, has a flux density of 0.20$\pm$0.01 Jy and was used as the reference point in the initial phase-referencing astrometry. The CLEAN component model of J1619+2247 was imported into AIPS, and the fringe-fitting was re-run to eliminate the source structure-dependent phase errors. In the second iteration of the phase-referencing calibration, we imaged R2 which shows a compact source of 19 mJy, and then applied the phase solutions derived from R2 to the data of AT2018cow and R1. Phase-referencing calibration using a closer calibrator in this way may further reduce the astrometric error.

\section{Results}

\subsection{Detection and flux density evolution}
The first session 2-hour electronic-EVN (e-EVN) observation at 1.67 GHz on 20 September 2018 ($t_d =$ 94.4 day) was aimed at an initial detection, characterizing its compactness and potential suitability for continued monitoring. The experiment successfully detected an unresolved compact source with a flux density of 0.86$\pm$0.04 mJy beam$^{-1}$, thus warranting continued monitoring \citep{2018ATel12067....1A}.  

The second session was carried out at 5 GHz on two sequential dates on 20 October 2018 ($t_d=$ 126.2 day) and 21 October 2018, each lasting 12 hr. The observations were made in disk-recording mode. The preliminary motivation was to resolve the source structure. This session offered the best resolution (synthesized beam size of 2.64 mas$\times$ 1.81 mas) and highest signal to noise ratio (SNR of $\approx$ 300) based on an image rms noise of 0.016 $\mu$Jy beam$^{-1}$ after self-calibration. An unresolved source was detected with a peak flux density of 4.80$\pm$0.04 mJy beam$^{-1}$, 5.6 times brighter than that in the first session. 

The third session was a 3-hour e-EVN observation at 5 GHz on 15 February 2019 ($t_d=$ 242.8 days). The source remains unresolved with a substantially decreased flux density of 0.38$\pm$0.03 mJy beam$^{-1}$. The resulting synthesized beam is 10.4 mas$\times$ 1.80 mas, substantially larger than other sessions due to the absence of long East-West baselines. 

\begin{figure}
\includegraphics[scale=1.]{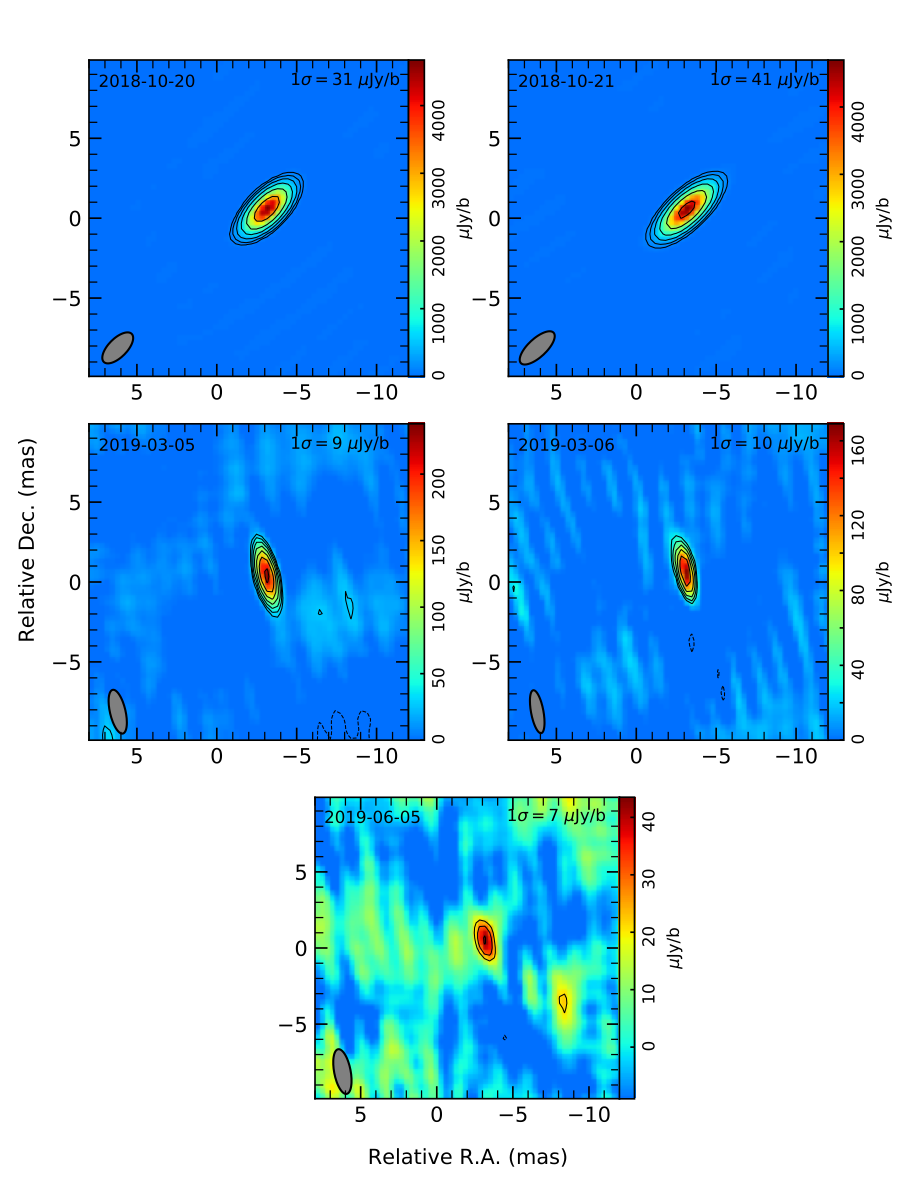}
\caption{High resolution 5 GHz VLBI images of AT2018cow during sessions 2, 4 and 5 indicating a compact, unresolved and fading transient. The observation date is in the top-left corner, the image noise threshold is in the top-right corner and the restoring beam shape (and size) are depicted in the bottom-left corner. The color scale represents the intensity (in units of $\mu$Jy beam$^{-1}$) with the lowest shown in blue and the highest shown in red.}
\label{fig1}
\end{figure}

The fourth session consisting of two 12-hr observations at 5 GHz on 5 and 6 March 2019 was aimed at monitoring the source structure or emission peak change. The source was detected at a peak flux density of 0.23$\pm$0.01 mJy beam$^{-1}$. A fifth session was a 12-hr observation at 5 GHz on 5 June 2019, again aimed at monitoring any structure or emission peak changes. Benefiting from the higher data rate of 2 Giga bits per second, this session has the highest sensitivity. The source was detected despite a vastly decreased flux density of 0.04$\pm$0.01 mJy beam$^{-1}$. The source images from Sessions 2 - 5 are presented in \autoref{fig1} and the imaging results are summarized in \autoref{table2} and include the observation day, observation frequency, the peak flux density and associated 1$\sigma$ error, the integrated flux density, the source structural parameters and the array used.

The 5 GHz light curve in \autoref{fig2} covers the early phase ($<$ 20 days), turnover (82.5 – 132.6 days) and later phase ($>$ 132.6 days). The light curve is fitted with a smoothly broken power law model based on a Markov Chain Monte Carlo (MCMC) method to obtain the temporal indices before and after the break, the `smoothness' of the break and the associated flux density and time at the break (see \autoref{fig2} and \autoref{fig3}, details in \autoref{app:lightcurve}), yielding a peak flux density of 5.6 mJy at a time of 102.1 days and a steep decline with an index $\leq-5.2$.

\subsection{Astrometry and proper motion}
The emission peak is determined by averaging the positions in individual sessions: $\alpha = 16^h16^m00^s.224169\pm0^s.000001$ and $\delta = 22^\circ16^{'}04^{''}.890539 \pm 0^{''}.000012$. The size of the morphologically unresolved source ranges from 0.068 mas in Session 2 to 0.35 mas in Session 5 based on Monte Carlo simulations. From the multi-epoch VLBI data spanning 260 days, the proper motion $1-\sigma$ upper limits are constrained to within  $\mu_{\alpha \cos\delta}$ = 0.070$\pm$0.049 mas yr$^{-1}$ and $\mu_\delta$ = 0.131$\pm$0.088 mas yr$^{-1}$ ($\leq$ 0.15$\pm$0.10 mas yr$^{-1}$ or 0.14$\pm$0.10 $c$). This conclusively rules out a relativistic jet, with a consequent intrinsically luminous expanding afterglow. 

\subsection{Source environmental properties}
An analytic model is used to fit the synchrotron emitting afterglow evolution \citep{2002ApJ...568..820G}, which assumes a self-similar adiabatically expanding ejecta interacting with the surrounding constant density or stratified medium \citep{1976PhFl...19.1130B,2013NewAR..57..141G}. Using the best-fitted VLBI size $\theta_A \leq$ 0.068 mas ($t_d=$ 126.2 days) and total energy $E=(0.01-0.32)\times 10^{52}$ erg \citep{2019ApJ...872...18M}, the number density is estimated to be $(0.12-8.00)\times 10^5$ cm$^{-3}$ at time $t_d=$ 22 days, consistent with the reported $3.0\times 10^5$ cm$^{-3}$ on the same date \citep{2019ApJ...871...73H}. Fractions of particle and magnetic field energy densities are  $\epsilon_e =$ 0.03 - 0.04 and $\epsilon_B \geq$ 0.33 respectively. As $\epsilon_B$ is larger than $\epsilon_e$ by an order of magnitude, this suggests a highly magnetized medium with $B \geq$ 0.84 G, calculated assuming that the magnetic field energy density in the co-moving frame is a fraction $\epsilon_B$ of the total energy density $\epsilon$, i.e. $B^2/(8 \pi)$ = $\epsilon_B \epsilon$, see \autoref{app:medium}. This independent estimate during the optically thin declining phase is consistent with the inferred $B \approx$ 6 G based on the expanding optically thick phase \citep{2019ApJ...871...73H}. Sub-mm polarimetric observations report the non-detection of linear polarization ($\leq$ 0.15 \%), attributable to Faraday depolarization in a dense and strongly magnetized medium \citep{2019ApJ...878L..25H}. The consistency of our independent measurements confirms the validity of the dense magnetized medium surrounding the transient and suggest that this environment persists well into the transient evolutionary phase.

\section{Discussion}
The expected number density is 1 cm$^{-3}$ and magnetic field strength is 1 $\mu$G in the tidal disruption scenario \citep{2019ApJ...878L..25H}. The inferred dense, magnetized medium  and the steeply declining flux density thus disfavor this scenario. The scenario of a failed explosion of a giant star which results in the direct collapse of the core to an accreting black hole is disfavored owing to the fall-back accretion powered light curve expected to decline with an index $\geq -2.4$ \citep{2018ApJ...857...95M}. A neutron star powered central engine is disfavored owing to the expected magnetic field strength in the shocked environment of  $\sim 10^4$ G  \cite[4 orders of magnitude higher than we infer, using the expressions for the equipartition magnetic field strength in \autoref{app:medium};][]{2019MNRAS.487.5618L}. We thus favor a scenario involving the successful supernova of a low-mass star which results in a newly formed magnetar powering the transient, as was speculated in literature \citep{2018ApJ...865L...3P,2019ApJ...871...73H,2019ApJ...878...34F} based on new, independent constraints from the VLBI observations during the late phase of transient evolution which substantiate this interpretation. The observations and implications in this context are further developed below. 

For a fast spinning magnetar of period 10 ms with a surface magnetic field strength $B_{15} = B/(10^{15} {\rm G})$, the rotational energy  $E_R \approx 2\times 10^{50}$ erg can potentially be lost to the surrounding medium through dipole radiation over a spin down timescale $t_s=$ 5.3 days. The ejecta from the supernovae can radiate this energy over a diffusion timescale $t_D=$ 3.8 days, resulting in a peak luminosity $L_p\approx $ 9.3$\times$10$^{44}$ erg s$^{-1}$, which then decays as $t^{-2}$ \citep{2010ApJ...717..245K}. The equipartition magnetic field strength in the surrounding medium \citep{2019MNRAS.487.5618L} ranges between $B_{eq} =$ 0.07 - 0.5 G at $t_d= t_p$ considering the cases of a wind-driven shock and the shock propagating through the low mass stellar ejecta. This is consistent within an order of magnitude of $B \geq$ 0.84 G, estimated above for the afterglow, supporting the scenario involving the magnetar driven winds causing the observed extremely luminous transient. 

The rapidly declining light curve ($\alpha_2 \leq -5.2$) challenges the presented scenario. Assuming (i) a rapid transition of the blast-wave to the Newtonian expanding phase, where the swept-up mass from the surrounding environment equals that in the initial ejecta \citep{2013NewAR..57..141G} ($\approx$ 0.1 - 0.4 $M_\odot$), and (ii) a density contrast of 0.01 between the material immediately in the vicinity of the blast-wave and that far away, the expected flux density can decline as $-$4.8 \citep{2000ApJ...541L..51K}, possibly explaining this observational challenge. 

An interesting consequence is the production of fast radio bursts (FRBs) from the magnetar interaction with the magnetized environment \citep{2016MNRAS.461.1498M,2017ApJ...841...14M,2019MNRAS.485.4091M}. This may provide a new understanding of FRBs, especially as three of these \cite[the repeating FRB 121102;][]{2017Natur.541...58C}, FRB 190523 \citep{2019Natur.572..352R} and FRB 180924 \citep{2019Sci...365..565B} have been precisely localised so far with a putative magnetar origin. The VLBI technique thus offers a promising, novel manner of understanding properties of transients (progenitor and evolution) as evidenced from the presented direct imaging observations of an FBOT. 

\acknowledgments

This work is supported by the SKA pre-research grant funded by the National Key R\& D Programme of China (2018YFA0404603) and the Chinese Academy of Sciences (CAS, 114231KYSB20170003). The VLBI data analysis was done on the China SKA Regional Centre prototype \citep{2019NatAs...3.1030A}. PM is supported by the CAS-PIFI (grant no. 2016PM024) post-doctoral fellowship and the NSFC Research Fund for International Young Scientists (grant no. 11650110438). T.A. thanks the grant from the Youth Innovation Promotion Association of CAS and from Gothenburg Centre for Advanced Studies in Science and Technology. The European VLBI Network is a joint facility of independent European, African, Asian, and North American radio astronomy institutes. Scientific results from data presented in this publication are derived from the following EVN project codes: RY007, EY033 and EM137. We thank the EVN PC for approving the trigger-of-opportunity observation, and the staff of the EVN and the observatories for carrying out the experiment. We thank the anonymous referee for quick suggestions that improved the content and presentation of our work. 

\begin{figure}
\includegraphics[scale=1.]{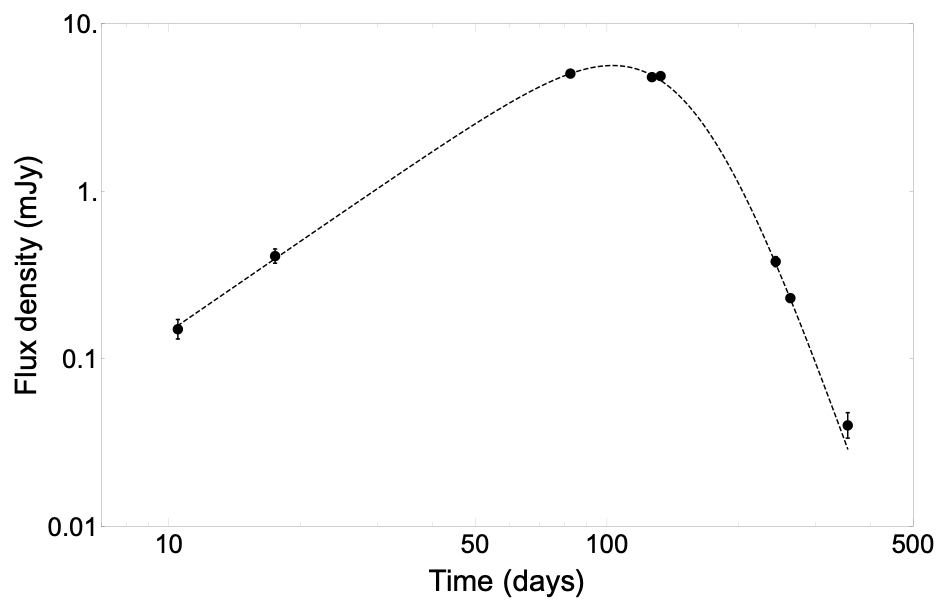}
\caption{MCMC fitting of the 5 GHz light curve of AT2018cow gives a steep declining phase slope of $-$7.5$\pm$1.1. A smoothly broken power law fit (see \autoref{app:lightcurve} for the light curve fitting methods adopted) gives a break flux density $F_{\nu,p}$ of 3.36$^{+0.60}_{-0.77}$ mJy, break time $t_p$ of 151.09$^{+14.75}_{-10.36}$ days, and temporal indices before and after the break $\alpha_1$ and $\alpha_2$ of 1.81$^{+0.07}_{-0.06}$ and $-$7.45$^{+0.90}_{-1.30}$ respectively, and smoothness parameter governing the peak turnover $s$ of 0.39$^{+0.10}_{-0.09}$. These correspond to a peak flux density of 5.6$\pm$1.4 mJy at a time of 102.1$\pm$13.1 days.}
\label{fig2}
\end{figure}

\begin{figure}
\includegraphics[scale=1.]{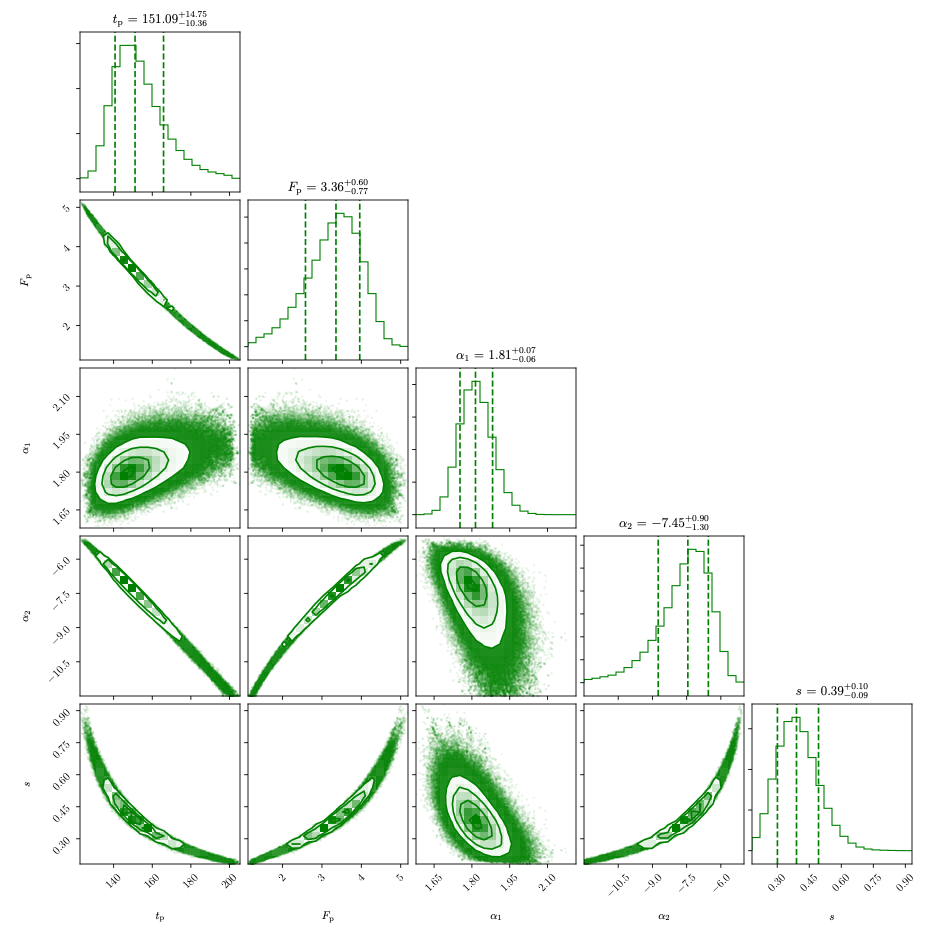}
\caption{Corner plot showing the MCMC fitting results. A smoothly broken power law fit (see \autoref{app:lightcurve} for the light curve fitting methods adopted) gives a break flux density $F_{\nu,p}$ of 3.36$^{+0.60}_{-0.77}$ mJy, break time $t_p$ of 151.09$^{+14.75}_{-10.36}$ days, and temporal indices before and after the break $\alpha_1$ and $\alpha_2$ of 1.81$^{+0.07}_{-0.06}$ and $-$7.45$^{+0.90}_{-1.30}$ respectively, and smoothness parameter governing the peak turnover $s$ of 0.39$^{+0.10}_{-0.09}$. The MCMC propagates both direct and co-variance based errors in these estimates and does not require prior knowledge of their underlying statistical distributions.}
\label{fig3}
\end{figure}

\begin{table}
\begin{center}
\begin{tabular}{lllllll} \hline
Session	& Project & Date   & $\nu_{\rm obs}$ & Time & BW & Radio telescopes \\
		& Code     & yymmdd & (GHz)           & (hrs)& (Mbps) & \\ \hline
1 & RY007A	& 180918 & 1.66	& 2.0 & 1024 & Jb2, Wb, Ef, Mc, On85, Tr, Sr, Cm, De, Kn	\\
2 & RY007B	& 181020 & 4.99 & 12.0 & 1024 & Jb2, Wb, Ef, Mc, On85, Ur, Tr, Ys, Hh, Sv, Zc, Bd, Ib, Ar, Km\\
  & RY007C	& 181021 & 4.99	& 12.0 & 1024 & Jb2, Wb, Ef, Mc, On85, Ur, Tr, Ys, Hh, Sv, Zc, Bd, Ib, Ar, Km\\
3 & EY033A	& 190215 & 4.93	& 3.0 & 2048 & Jb2, Ef, Nt, Mc, On85, Tr, Ys, Hh, Ib\\
4 & EY033B & 190305	& 4.99 & 12.0 &1024	& Jb2, Wb, Ef, Mc, Nt, On85, T6, Ur, Tr, Ys, Hh, Sv, Zc, Bd, Ib\\
  & EY033C & 190306 & 4.99 & 12.0 & 1024 & Jb2, Wb, Ef, Mc, Nt, On85, T6, Ur, Tr, Ys, Hh, Sv, Zc, Bd, Ib\\
5 & EM137 & 190605 & 4.93 & 12.0 & 2048 & Jb2, Wb, Ef, Mc, Nt, On85, T6, Tr, Ys, Hh, Sv, Zc, Bd, Ir\\ \hline
\end{tabular}
\caption{The experiment setup for the EVN observations of AT2018cow. Columns give (1) Session ID, (2) EVN project code, (3–4) observing date and frequency, (5–6) total observing time and recording bandwidth, and (7) a total of 21 participating radio telescopes: Ar (Arecibo 300m, USA), Bd (Badary 32m, Russia),  Cm (Cambridge 32m, UK), De (Defford 25m, UK), Ef (Effelsberg 100m, Germany), Hh (Hartebeesthoek 26m, South Africa), Ib (Irbene 32m, Latvia), Ir (Irbene 16m, Latvia), Jb2 (Lovell 76m, UK), Km (Kunming 40m, China), Kn (Knockin 25m, UK), Mc (Medicina  25m, Italy), On85 (Onsala 25m, Sweden), Sr (Sardinia 64m, Italy), Sv (Svetloe 32m, Russia), T6 (Tianma 65m, China), Tr (Torun 32m, Poland), Ur (Urumqi 26m, China), Wb (Westerbork 25m, Netherlands), Ys (Yebes 40m, Spain), Zc (Zelenchukskaya 32m, Russia).}
\label{table1}
\end{center}
\end{table}

\begin{table}
\begin{center}
\begin{tabular}{lllllllllll} \hline
Session	& MJD & $t_{post}$ & $\nu_{obs}$ & $S_{peak}$ & $\sigma_{map}$ & $S_{int}$ & $\phi_{maj}$ & $\phi_{min}$ & $\phi_{pa}$ & Array \\
 & (day) & (day) & (GHz) & (mJy beam$^{-1}$) & (mJy beam$^{-1}$) & (mJy) & (mas) & (mas) & (deg) & \\ 
 & (1)   & (2)   & (3)   & (4)               & (5)               & (6)   & (7)   & (8)   & (9)   & (10) \\ \hline
1	&58379.8	&94.4	&1.67	&0.86	&0.040	&0.89	&48.3	&15.0	&+45.1	&e-EVN\\
2	&58411.6	&126.2	&4.99	&4.79	&0.031	&4.67	&2.64	&1.81	&+08.9	&Full EVN\\
	&58412.6	&127.2	&4.99  	&4.80	&0.041	&4.72	&2.59	&2.08	&+09.3	&Full EVN\\
3	&58528.2	&242.8	&4.93	&0.38	&0.026	&0.37	&10.4	&1.80	&+78.8	&e-EVN\\
4	&58547.2	&261.8	&4.99	&0.27	&0.009	&0.24	&2.81	&0.97	&+13.1	&Full EVN\\
	&58548.2	&262.8	&4.99	&0.19	&0.010	&0.17	&2.73	&0.78	&+10.6	&Full EVN\\
5	&58640.0	&354.6	&4.93	&0.043	&0.007	&0.043	&2.97	&1.04	&+11.4	&Full EVN\\ \hline
\end{tabular}
\caption{The EVN phase-referencing imaging results of AT2018cow. Columns give (1) Modified Julian Date (MJD), (2) post-explosion day (reference time: MJD 58285.4), (3) central observing frequency, (4) peak brightness in mJy beam$^{-1}$ in the dirty map, (5) off-source image noise level, (6) total flux density derived by fitting the visibility data to a point source model, (7--9) the major axis, the minor axis and the position angle of the elliptical Gaussian beam synthesized with natural weighting, and (10) network configuration. Note that the nominal systematic errors due to visibility amplitude calibration for $S_{peak}$ and $S_{int}$ are $\sim$ 5 \%.}
\label{table2}
\end{center}
\end{table}

\appendix

\section{VLBI Observational results}
\label{app:structure}
A morphologically compact unresolved mas-scale radio source is unambiguously detected in AT2018cow in all epochs. The flux densities of AT2018cow and the related image parameters are listed in \autoref{table2}. We should note that $\sigma_{\rm map}$ only represents the random fluctuations in the image, i.e., the rms noise. An additional 5\% systematic error of the flux density measurements is included in the calibration of the visibility amplitude. 

There is no hint of any secondary component or extension in the residual images at any session with diverse $(u,v)$ coverages and dynamic ranges. The total flux density fitted with a point source model is consistent with the peak flux density in the dirty map, supporting the detection of an extremely compact source. In order to estimate the source size in such an unresolvable case, we performed one thousand Monte Carlo simulations: in each simulation, we first subtracted the observed source AT2018cow from the visibility data, then we added a point source with the same flux density as that of AT2018cow at a random position, next we fitted the fake source with a circular Gaussian model. The sizes at the cumulative probabilities of 68.3\%, 95.4\%, 99.7\% are reported in column 6 in \autoref{tableA1}, and the minimum brightness temperature based on the size constraint at 99.7\% in column 7. Sessions 1 and 3 resulted in larger beam sizes due to missing long baselines, therefore they are not included in this estimate.

\section{Astrometry and proper motion}
\label{app:astrometry}
R2 shows a compact structure (angular size $\approx$ 0.3 mas) and a flux density of 18.9 $\pm$ 1.0 mJy at 5 GHz and 15.6 $\pm$ 0.8 mJy at 1.67 GHz. We measured its position: $\alpha$(J2000) = 16$^h$16$^m$40$^s$.8140683, $\delta$(J2000) = 22$^\circ$18$^{'}$56$^{''}$.410927 (J2000) from the first full-EVN observation in Session 2. Comparing the peak positions in all sessions results in a formal positional uncertainty of only a few micro-arcsecond ($\mu$as). Then we derived a weighted average position of AT2018cow: $\alpha$(J2000) = 16$^h$16$^m$00$^s$.224169$\pm$0$^s$.000001 and $\delta$(J2000) = 22$^\circ$16$^{'}$04$^{''}$.890539 $\pm$ 0$^{''}$.000012. As the primary calibrator J1619+2247 comprises an elongated jet with a flux density of 0.41$\pm$0.02 Jy, that may give rise to a systematic positional uncertainty up to 0.8 mas. In order to avoid this systematic uncertainty of the absolute position, we focus on the differential astrometry, i.e., the relative motion between R2 and AT2018cow, in the following discussion. Only the formal position uncertainty is taken into accounted. 

R1 is used as a checker source to evaluate the phase-referencing quality. R1 is successfully detected in the dirty image without any self-calibration with an SNR $\approx$ 100. It has a flux density of 15.1 $\pm$ 0.8 mJy at 5 GHz and 23.8 $\pm$ 1.2 mJy at 1.67 GHz, respectively. The source shows a resolved structure consisting of two components. The brighter and more compact component is most likely the core. In Session 2, we found some phase errors on the baselines associated with Ibbene and Arecibo. Thus, the two data of these two stations were excluded from the astrometric study. With respect to R2, the statistical average position of R1 is: $\alpha$(J2000) = 16$^h$15$^m$41$^s$.6911175$\pm$0.0000015, $\delta$(J2000) = 22$^\circ$16$^{'}$28$^{''}$.222084$\pm$0.000025. Since both AT2018cow and R1 have similar angular distances, 9.8 and 13.9 arcmin, to R2 respectively, and their data were calibrated in the same way, their systematic position errors are expected to be roughly the same. Thus, the astrometric accuracy is mainly associated with the position accuracy of R2.

\begin{figure}
\includegraphics[scale=.4]{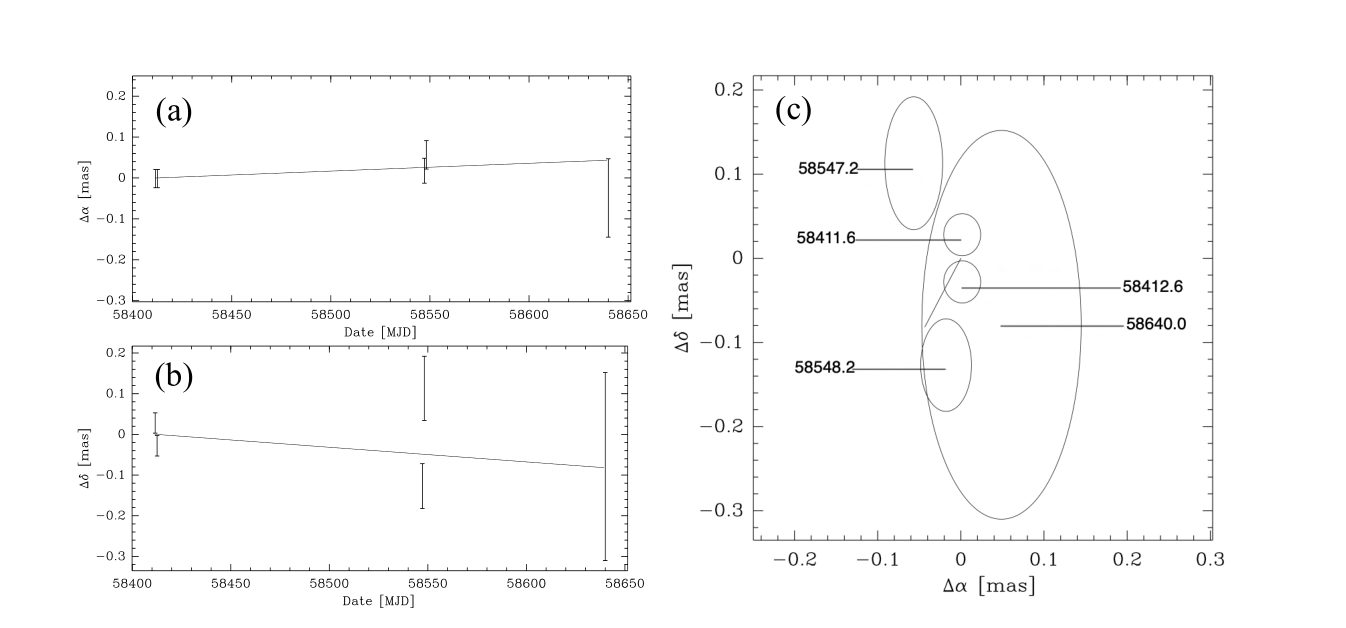}
\caption{No significant proper motion of AT2018cow. (a) Plot of the position offset in Right Ascension versus time. (b) Plot of the position offset in Declination versus time. (c) The scatter plot of the measured position offsets. All the error bars and ellipses represent 1$\sigma$ uncertainties and the epochs corresponding to the appropriate session are listed in MJD. The straight lines in panels a and b represent the proper motion searching results: $\mu_{\alpha \cos \delta}$ = 0.070 $\pm$ 0.049 mas yr$^{-1}$ and $\mu_\delta$ = $–$0.131 $\pm$ 0.088 mas yr$^{-1}$.}
\label{figA1}
\end{figure}

The \autoref{figA1} shows the astrometric results from the full EVN observations (Sessions 2, 4 and 5) involving more telescopes. As the transient faded significantly in the Sessions 4 and 5, the error bars and ellipses became much bigger. Using the software PMPAR\footnote{https://github.com/walterfb/pmpar}, we estimated a proper motion of $\mu_{\alpha \cos \delta}$ = 0.070 $\pm$ 0.049 mas yr$^{-1}$ and $\mu_\delta$ = $–$0.131 $\pm$ 0.088 mas yr$^{-1}$. The fitted proper motions only account for 1.5$\sigma$, and yet suggest no significant proper motion in AT2018cow. 

\section{Light curve fitting}
\label{app:lightcurve}
The 5 GHz light curve in \autoref{fig2} is compiled from reported and measured flux densities covering the early phase ($<$ 20 days), turnover (82.5 – 132.6 days; including our Session 2) and later phase ($\approx$ 132.6 days; including our Sessions 3-5). The light curve is fitted with a smoothly broken power law model $F_\nu=2^{1/s} F_{\nu,p}  \left((t/t_p)^{-s \alpha_1}+(t/t_p)^{-s \alpha_2}\right)^{-1/s}$, where $F_{\nu,p}$ is the break flux density (mJy), $t_p$ is the break time (day), $\alpha_1$ and $\alpha_2$ are the temporal indices before and after the break respectively, and $s$ is a smoothness parameter governing the peak turnover. A least squares fitting gives ($\alpha_1$,$\alpha_2$,$s$) = (1.7$\pm$0.1,$-$5.2$\pm$0.2,0.9$\pm$0.1). With a ($F_{\nu,p}$,$t_p$) = (5.0$\pm$0.1 mJy, 124.4$\pm$1.4 days), we estimate a peak flux density of 5.9$\pm$0.1 mJy at a time of 103.3$\pm$1.9 days. 

The slope $\alpha_2$ = $-$5.2$\pm$0.2 derived from the least squares fitting is much steeper than other known transients. We then employed a Markov Chain Monte Carlo (MCMC) method \cite[emcee Python package;][]{2013PASP..125..306F} to better constrain these parameters without any prior knowledge of their underlying statistical distributions. The MCMC results are shown in the corner plot Figure 3 and are ($\alpha_1$,$\alpha_2$,$s$) = (1.81$^{+0.07}_{-0.06}$,$-$7.45$^{+0.90}_{-1.30}$,0.39$^{+0.10}_{-0.09}$). With a ($F_{\nu,p}$,$t_p$) = (3.36$^{+0.60}_{-0.77}$ mJy, 151.09$^{+14.75}_{-10.36}$ days), we estimate a peak flux density of 5.6$\pm$1.4 mJy at a time of 102.1$\pm$13.1 days. As MCMC results in an even steeper slope of $\alpha_2$ =$-$7.45$^{+0.90}_{-1.30}$, the true slope is taken to be $\leq -$5.2. The fitting results are summarized in \autoref{tableA2}.

\section{Afterglow spectral evolution}
\label{app:afterglow}
The source size, flux density and light curve fitting parameters are used in an analytic model depicting the synchrotron emitting afterglow evolution \citep{2002ApJ...568..820G}. The model assumes a self-similar adiabatically expanding ejecta interacting with the surrounding medium \citep{1976PhFl...19.1130B,2013NewAR..57..141G} in two scenarios. The first (Model 1) is characterized by a medium of constant density $n_0$. The second (Model 2) is characterized by a stratified wind-like medium with a density profile $A_\ast r^{-2}$ , where $A_\ast$ is an effective measure of the mass density, and $r$ is the radial distance from the central core. The current data are not able to distinguish between these two models, therefore we make the subsequent calculations in both scenarios.

Associated emission frequencies mark distinctive breaks in the evolving synchrotron spectrum, corresponding to transitions from fast-cooling ($\nu_c< \nu_m$) to slow-cooling ($\nu_m < \nu_a$) phase. $\nu_a$ is the synchrotron self-absorption frequency characterising a surface whose optical depth to synchrotron self-absorption is unity; $\nu_m$ is the synchrotron frequency emitted by the power law distributed electrons; and $\nu_c$ is the synchrotron frequency of an electron which cools over the dynamic timescale. 

\subsection{Model 1 (constant density ISM)}
The synchrotron frequency $\nu_m$ and synchrotron self-absorption frequency $\nu_a$ are,
\begin{align}
\nu_m &= (1.58 \times 10^{13}~{\rm Hz}) E^{1/2}_{52} \epsilon^2_e \epsilon^{1/2}_B t^{-3/2}_d \\ \nonumber
\nu_a &= (4.67 \times 10^{11}~{\rm Hz}) E^{0.24}_{52} n^{0.13}_0 \epsilon^{0.46}_e \epsilon^{0.12}_B t^{-0.73}_d.
\end{align}
With $\phi_m = (\nu/\nu_m)$, the corresponding composite spectrum as a function of the observation frequency $\nu$ (in GHz) is given by,
\begin{align}
F_\nu = &F_{\nu,{\rm max}} \left(\phi^2_m  e^{-8.22 \phi^{2/3}_m}+\phi^{5/2}_m \right) \left(1+(\nu/\nu_a)^{2.14}\right)^{-1.82}\\ \nonumber
&F_{\nu,{\rm max}} = (1.89\times 10^{10}~{\rm mJy}) E^{3/2}_{52} n^{-1/2}_0 \epsilon^5_e \epsilon_B t^{-5/2}_d.
\end{align}

\subsection{Model 2 (stratified ISM)}
The synchrotron frequency $\nu_m$ and synchrotron self-absorption frequency $\nu_a$ are,
\begin{align}
\nu_m &= (2.54\times 10^{13}~{\rm Hz}) E^{1/2}_{52} \epsilon^2_e \epsilon^{1/2}_B t^{-3/2}_d \\ \nonumber
\nu_a &= (6.73 \times 10^{11}~{\rm Hz}) E^{0.12}_{52} A^{0.26}_{\ast} \epsilon^{0.46}_e \epsilon^{0.12}_B t^{-0.86}_d.
\end{align}

The corresponding composite spectrum as a function of the observation frequency $\nu$ (in GHz) is given by,
\begin{align}
F_\nu = &F_{\nu,{\rm max}} \left(\phi^2_m  e^{-8.56 \phi^{2/3}_m}+\phi^{5/2}_m \right)  \left(1+(\nu/\nu_a)^{2.31}\right)^{-2.03} \\ \nonumber
&F_{\nu,{\rm max}} = (1.54\times 10^{10}~{\rm mJy}) E^{2}_{52} A^{-1}_{\ast} \epsilon^5_e \epsilon_B t^{-2}_d.
\end{align}

Model parameters include the explosion energy $E_{52} = E/(10^{52}~{\rm erg~s^{-1}})$, number density of the surrounding medium ($n_0$ in Model 1, or $A_\ast r^{-2}$  in Model 2), and fractions of the shock energy density in the particles $\epsilon_e$ and magnetic fields $\epsilon_B$. The source angular size is related to $E_{52}$, $n_0$ or $A_\ast$, $t_d$ by $\theta_{A1} = (0.83~{\rm mas}) E^{1/4}_{52} n^{-1/4}_0 t^{1/4}_d$ (Model 1) and $\theta_{A2} = (0.32~{\rm mas}) E^{1/2}_{52} A^{-1/2}_\ast t^{1/2}_d$ (Model 2). Using the best-fitted VLBI size $\theta_A {\rm(epoch~2)} \leq 0.07$ mas ($t_d$ = 126.2 days) and $E_{52}$ = 0.01 - 0.32 \citep{2019ApJ...872...18M}, we obtain $n_0$ = (2.5-80.0) $\times 10^4$ cm$^{-3}$ for Model 1, and $A_\ast$ = 26.3 - 842.1 for Model 2 corresponding to a number density $n$ =(1.2-39.0)$\times 10^4$ cm$^{-3}$ at an observation time $t_d$ = 22 day. These estimates are consistent with the reported $3.0 \times 10^5$ cm$^{-3}$ on $t_d$ = 22 days \citep{2019ApJ...871...73H}, confirming a dense surrounding medium in AT2018cow. 

\section{Physical conditions in the surrounding medium and implications on the progenitor}
\label{app:medium}
The model flux density $F_\nu (E_{52},n_0~{\rm or}~A_\ast,\epsilon_e,\epsilon_B,t_d)$ when subject to the conditions $F_\nu (t_d = t_p )= F_{\nu,p}$ and $F^{'}_\nu (t_d = t_p)= 0$ yields  $\epsilon_e$ = 0.03 - 0.04 and $\epsilon_B \geq$ 0.33, suggesting a magnetically dominated medium. For the magnetic field energy density (in the co-moving frame) being a fraction $\epsilon_B$ of the total energy density $\epsilon = 4 \gamma^2 n m_p c^2$, i.e. $B^2/(8 \pi)= \epsilon_B \epsilon$, this corresponds to $B_1 = (1.43~{\rm G}) E^{1/8}_{52} n^{3/8}_0 \epsilon^{1/2}_B t^{-3/8}_d$ (Model 1) and $B_2 =(1.45~{\rm G}) E^{-1/4}_{52} A^{3/4}_\ast \epsilon^{1/2}_B t^{-3/4}_d$ (Model 2), based on $\gamma_1 = 3.67~E^{1/8}_{52} n^{-1/8}_0 t^{-3/8}_d$ (Model 1) and $\gamma_2 = 3.72~E^{1/4}_{52} A^{-1/8}_\ast t^{-1/4}_d$ (Model 2). Using the above parameters and assuming $t_d$ as the peak time, we get $B_1 \geq$ 3.42 G  (Model 1) and $B_2 \geq$ 0.84 G (Model 2). 

The magnetar scenario is now explored further to estimate the energetics \citep{2010ApJ...717..245K} and magnetic field strength in the medium \citep{2019MNRAS.487.5618L}. The moment of inertia of the magnetar $I=(2/5) M R^2 = 1.1 \times 10^{45}$ g cm$^2$ assuming a mass $M= 1.4 M_\odot$ and radius $R$ = 10 km. With magnetar spin period $P_{-2} = P/(10~{\rm ms})$, the rotational energy available is $E_R = 2 \pi^2 I P^{-2}$ = $(2.2 \times 10^{50}~{\rm erg}) P^{-2}_{-2}$. For a magnetar with a surface magnetic field strength $B_{15} = B/(10^{15}~{\rm G})$, this energy can potentially be lost to the surrounding medium through dipole radiation over the spin down timescale $t_s  = 3 I c^3 P^2 (2 \pi^2  B^2  R^2)^{-1}$  = $(5.3~{\rm days}) B^{-2}_{15} P^2_{-2}$. The adiabatically expanding ejecta slows down upon emitting a bulk of this energy over the diffusion timescale $t_D = (\kappa M_{ej}/(4 v_{ej} c))^{1/2}$ = 3.8 day assuming a diffusion coefficient $\kappa$ = 0.2 g$^{-1}$ cm$^2$, an ejecta mass $M_{ej}$ = 0.3 $M_\odot$ and a velocity $v_{ej}$ = 0.1 $c$. This results in a peak luminosity $L_p \approx E_R t_s/t^2_D$  = $(9.3 \times 10^{44}~{\rm erg~s^{-1}}) B^{-2}_{15}$. 

The luminosity then decays as a power law $L = L_p (1+t/t_s)^{-2}$ \citep{2019ApJ...872...18M} based on which the equipartition magnetic field strength in the surrounding medium  for a wind-driven shock \citep{2019MNRAS.487.5618L} is $B_{eq} = (4.9 \times 10^3~{\rm G}) B^{-1/2}_{15} t^{-5/4}_d (1+t_d/t_s)^{-1}$ for the shock propagating through the ejecta (assuming an ejecta mass $M_{ej}$ = 0.3 $M_\odot$ and a velocity $v_{ej}$ = 0.1 $c$), and $B_{eq} = (2.1 \times 10^2-1.2 \times 10^3~{\rm G}) t^{-1}_d (1+t_d/t_s)^{-1}$ for the shock propagating through the pre-explosion wind. In the latter case, we use $\dot{M}/v_w = (1.66\times 10^{14} - 5.31\times 10^{15})$ g cm$^{-1}$ based on the inferred $A_\ast$ = 26.3 - 842.1 constrained from the VLBI size of the source. For $t_d$ being the peak time, $B_{eq}$ = 0.07-0.5 G in both cases, consistent to within an order of magnitude of the estimate made above of $B_2$ (and $B_1$)$\geq$ 0.84 G, providing support to the scenario involving the magnetar driven winds causing the observed extremely luminous transient. For the case of a regular neutron star with $B$ = 10$^9$ G, the magnetic field strength in the shock can be as high as $B_{eq}$ = 1.2$\times$10$^4$ G, much higher than the estimate made here, providing considerably less support to scenarios involving a neutron star such as that in a common envelope.


\begin{table}
\begin{center}
\begin{tabular}{lllllll} \hline
MJD	& Right Ascension & $\sigma_\alpha$ & Declination & $\sigma_\delta$ & $\theta_{sim}$ & $T_{b,min}$ \\ 
(day) & $\alpha$(J2000) & (mas) & $\delta$(J2000) & (mas) & (mas) & (K)\\ \hline
58411.6	&16$^h$16$^m$00$^s$.2241686	&0.0017	&22$^\circ$16$^{'}$04$^{''}$.890567	&0.0025	&0.068, 0.12, 0.14	&1.2$\times$10$^{10}$\\
58412.6	&16$^h$16$^m$00$^s$.2241686	&0.0021	&22$^\circ$16$^{'}$04$^{''}$.890511	&0.0026	&0.064, 0.13, 0.16	&9.2$\times$10$^9$\\
58547.2	&16$^h$16$^m$00$^s$.2241700	&0.021	&22$^\circ$16$^{'}$04$^{''}$.890412	&0.049	&0.19, 0.31, 0.35	&9.7$\times$10$^7$\\
58548.2	&16$^h$16$^m$00$^s$.2241728	&0.026	&22$^\circ$16$^{'}$04$^{''}$.890652	&0.075	&0.22, 0.39, 0.47	&3.8$\times$10$^7$\\
58640.0	&16$^h$16$^m$00$^s$.2241652	&0.093	&22$^\circ$16$^{'}$04$^{''}$.890460	&0.23	&0.35, 0.76, 1.42	&1.1$\times$10$^6$\\ \hline
\end{tabular}
\caption{The astrometric results and the constraints on the size of AT2018cow. Columns give (1) MJD, (2--3) Right Ascension and 1$\sigma$ formal uncertainty (systematic uncertainty: 0.022 mas), (4–5) Declination and 1$\sigma$ formal uncertainty (systematic uncertainty: 0.025 mas). (6) sizes at the cumulative probabilities of 68.3\%, 95.4\%, 99.7\% from Monte Carlo simulation of a fake point source, (7) The lower limit of the brightness temperature.}
\label{tableA1}
\end{center}
\end{table}

\begin{table}
\begin{center}
\begin{tabular}{lllllll} \hline
Fitting method & $F_{\nu,p}$ & $t_p$ & $\alpha_1$ & $\alpha_2$ & $s$ \\
               & (mJy)       & (days) &           &            &     \\ \hline
Chi-square &5.0$\pm$0.1 & 124.4$\pm$1.4 & 1.7$\pm$0.1 & $-$5.2$\pm$0.2 & 0.9$\pm$0.1\\
MCMC & 3.4$\pm$0.7 & 151.1$\pm$12.8 & 1.8$\pm$0.1 & $-$7.5$\pm$1.1 & 0.4$\pm$0.1\\ \hline
\end{tabular}
\caption{The model fitting parameters for the 5 GHz light curve. Column (1) is the method used to fit the light curve, columns (2-6) are the results of the fitting along with their corresponding 1$\sigma$ errors.}
\label{tableA2}
\end{center}
\end{table}

\end{document}